\documentstyle[preprint,aps,eqsecnum]{revtex}
\begin{document}
\setlength{\baselineskip}{22pt}
\parindent=20pt\parskip=4pt plus 1 pt
\tightenlines
\title
{{\hspace*{12 cm} DFF 361--7--00}\hfill\break
\vskip 2truecm
{Density fluctuations and multifragmentation \\
of nuclear matter}}

\author{ F. Matera and A. Dellafiore\\
{\small\it Dipartimento di Fisica, Universit\`a degli Studi di Firenze, }\\
{\small\it and Istituto Nazionale di Fisica Nucleare, Sezione di
Firenze,}\\
{\small\it L.go E. Fermi 2, I-50125, Firenze, Italy} \\}
\maketitle
\begin{abstract}
 The density fluctuations of nuclear matter are studied within a
mean--field model in which fluctuations are generated by an external
stochastic field. The constraints imposed on the random force by the
fluctuation-dissipation theorem are analyzed.  It is shown that
in the proximity of the borders of the spinodal region the assumption
of a withe--noise stochastic field can be reliably used.
The domain distribution of the liquid phase in the spinodal
decomposition of nuclear matter is derived. The related distribution
of fragment sizes compares favourably with the experimental fragment
distribution observed in heavy ion collisions.
\end{abstract}
\vskip 1.5cm
PACS number(s): 21.65.+f, 24.60.Ky, 25.70.Pq, 21.60.Jz 

\vskip 2.5cm

\section{Introduction}

Semiclassical kinetic equations for the one--body phase--space
density provide a powerful tool for
studying the dynamics of complex processes occurring in heavy
ion collisions \cite{Ber88,Schu89,Bon94}. However, these
equations in their original version give a deterministic
description for the evolution of the one--body phase--space
density and their solution represents the mean value of this density
at each time. Thus, they are not able to account for phenomena such
as the nuclear multifragmentation observed in heavy ion collisions.
In this process fluctuations about the mean phase--space density
are believed to play an essential role (for a review on nuclear
multifragmentaion
see, e.g., Ref.\cite{Tam98}). In the last decade, to remedy for
this drawback, an extension of the transport theory
has been proposed \cite{Ay90,Ran90,Col94}. This approach,
that incorporates a fluctuating stochastic term into the kinetic
equation, is usually known as the Boltzmann--Langevin equation,
and  was originally applied to the treatment of hydrodynamic
fluctuations  in the theory of classical fluids \cite{Bix69}.
In Refs.\cite{Ay90,Ran90,Col94} collisions
between nucleons in the nuclear medium are regarded as random
processes and the diffusion coefficient of the Langevin (fluctuating)
term is ultimately related to the amplitude of the
nucleon--nucleon scattering. The last step is a particular
case of the fluctuation--dissipation theorem. More recently,
a new method to take into account fluctuations has also been
proposed \cite{ColA98}. In the approach of Ref.\cite{ColA98} the
statistical fluctuations of the one--body phase--space density
are directly introduced by assuming local thermodynamic equilibrium.
The white--noise nature of the stochastic term is the basic 
assumption, generally shared by all works on this subject. The 
authors of Ref.\cite{Ay94}, instead, have introduced 
an extension of the Boltzmann--Langevin theory  by including 
a coloured--noise term in the stochastic force, the occurrence 
of such a term has been ascribed to the finite nucleon--nucleon 
collision time. Actual applications to nuclear problems of this 
interesting approach still have to be made.
\par
In the present paper, we study  the density
fluctuations and their time evolution by introducing a
self--consistent stochastic field acting on the constituents of
the system. The self--consistency condition is provided by
the fluctuation--dissipation theorem. The evolution of the
fluctuations is treated within a linear approximation for the
stochastic field. For simplicity we consider an infinite 
homogeneous system. First, without introducing any particular
assumption, we prove that a withe--noise stochastic field
cannot satisfy the self--consistency condition in general.
Then, with reference to infinite nuclear matter, and within a
collisionless mean--field approximation, we specify the 
particular conditions under which the  withe--noise assumption
for the stochastic field can be retained. These
conditions are fulfilled for values of density and
temperature lying in the proximity of the boundary
of the spinodal region in the phase diagram, both
inside and outside this region. Thus we consider nuclear
matter in this physical situation and
are able to solve the stochastic equation for the density
fluctuations in a closed form.
\par
With respect to the previous works on this  
subject \cite{Ay90,Ran90,Col94}, here we consider
a different source of fluctuations: Landau damping. This source
is present even when collisions have a negligible role
in the evolution of the system.
\par
In Sec. II we propose a procedure to determine the
structure of domains formed within the system during a spinodal
decomposition. If the fragmentation phenomenon observed
in heavy ion collisions can be ascribed to a spinodal
decomposition of the bulk of nuclei \cite{Ber83},
we are allowed to identify the pattern according to which nuclear
matter is decomposing, with the fragmentation pattern, and can compare
the results of our nuclear matter calculations with the fragment
distribution observed in heavy ion collisions. This comparison is made
in Sec. III. Finally in Sec. IV a brief summary and conclusions are
given.
\par
Many papers, both theoretical and experimental, have
been devoted to the multifragmentation problem. Here we mention 
only a few theoretical ones representing different approaches.
In the statistical models of Refs.\cite{Gross,Bond95}, a complete 
statistical equilibrium of all degrees of freedom is assumed in a 
freeze--out volume and the various exit channels are sorted 
according to their statistical weight in the microcanonical ensemble.
In Ref.\cite{Mor97} instead nuclear multifragmentation has been
described in terms of "reducibility" and "thermal scaling".
This means that fragments are emitted practically independently
of each other and the one--fragment probability is given by
a Boltzmann factor. In the dynamical models of Refs.
\cite{Ay90,Ran90,Col94} clusters are constructed from the one--body
phase--space density governed by the Boltzmann-Langevin equation
\cite{Col96,Col97,ColB98}. In order to take into account the quantal
nature of the system and the requirement of antisymmetrization,
the Quantum Molecular Dynamics model \cite{Aiche} and the more
sophisticated Fermionic (Antisymmetrized) Molecular Dynamics model
\cite{Feld97,Ono96} have been developed. 
In addition, percolation \cite{Bauer} and lattice--gas models
\cite{Cam97,DGu97} have also been introduced. These models
are particularly suitable to deal with the critical phenomena
which can be expected to occur in multifragmantation. To conclude
this non--exhaustive survey, we mention the calculations of 
Ref.\cite{Bra96} that are based on the classical 
nucleation theory.
\par
Even if it should eventually turn out that multifragmentation 
must be ascribed to very complicated processes, we think 
that in any case  our present approach could give some insight 
into the underlying mechanism.

\section{Formalism}
\subsection{White--noise assumption}

The mean--field approximation allows us to obtain a 
self--consistent equation for the time evolution of the 
one--body density. We assume that the time scale of the terms  
neglected in the mean--field approximation is shorter than 
the characteristic times of mean--field dynamics. In order to 
take into account thermodynamic fluctuations, quantum effects 
and short--range correlations, we add to the mean--field 
a stochastic term similar to the random force in the Langevin 
equation. We assume that this additional field is a gaussian white 
noise with vanishing mean. In this case the time--evolution of 
the density is a markovian process. We will determine the conditions 
in which the white--noise assumption can be considered reasonable.

The additional stochastic mean field will induce density 
fluctuations with respect to the mean density.
To be more specific, we assume that at the time $t=0$ in 
the system is present a density fluctuation
$\delta\varrho({\bf r},t=0)$, with $\delta\varrho({\bf r},t)=
\varrho({\bf r},t)-\varrho_0$, 
( $\varrho_0$ is the density of the reference homogeneous state
i.e the state towards which the system relaxes ).
Within a linear approximation for the stochastic mean--field
the Fourier coefficients of $\delta\varrho({\bf r},t)$
for $t>0$ are given by ( see for example Ref.\cite{Reic}, Sec. 15~I )
\begin{equation}
\delta\varrho_{\bf k}(t)=\,\delta\varrho_{\bf k}(t=0)-
\frac{\delta\varrho_{\bf k}(t=0)}
{D_{k}(\omega=0)}\,\int_{0}^{t}D_{k}(t-t^\prime)\,dt^\prime\,+
\int_{0}^{t}D_{k}(t-t^\prime)B_{\bf k}(t^\prime)dW_{\bf
k}(t^\prime)\,,
\label{wiener}
\end{equation}
where $D_{k}(t-t^\prime)$ is the response function of the
nuclear medium and $D_{k}(\omega)$  its time Fourier transform.
For symmetry reasons $D_{k}(t-t^\prime)$ and its Fourier transform
depend only on the magnitude of the wawe vector. In the second 
integral $B_{\bf k}(t^\prime)dW_{\bf k}(t^\prime)$ gives the 
contribution of the stochastic field in the interval $dt^\prime$. 
The real and imaginary parts of the Fourier coefficients 
$W_{\bf k}(t^\prime)$ are indipendent components of a multivariate 
Wiener process \cite{Gard}. The fact that the stochastic 
field is real requires $B^{*}_{\bf k}(t)=B_{-\bf k}(t)$ and 
$W^{*}_{\bf k}(t)=W_{-\bf k}(t)$. 
\par
The stochastic part of the mean field is completely determined 
once the coefficients $B_{\bf k}(t)$ are known. In order to gain
information about these coefficients we concentrate on
the correlations of density fluctuations at equilibrium.
Due to the independence of the components of the multivariate
Wiener process $W_{\bf k}(t)$, only the terms with
${\bf k}^\prime=-{\bf k}$ survive. Within a linear approximation,
these correlations can be expressed by means of the same
quantities that appear in  Eq.(\ref{wiener}). This does, in a sense,
correspond to the Onsanger hypothesis about the decay of deviations
from equilibrium \cite{Reic}. The equation for the equilibrium
fluctuations is obtained from Eq.(\ref{wiener}) by moving the 
initial time to $-\infty$, without including any particular 
condition at finite times. Then the correlations are given by 
the equation ( the brackets denote ensemble averaging )
\begin{equation}
<\delta\varrho_{\bf k}(t)\delta\varrho_{-\bf k}(t^\prime)>=\,
\int_{-\infty}^{min(t,t^\prime)}dt_{1}\,D_{k}(t-t_1)D_{k}(t^\prime-t_1)\,
B_{\bf k}(t_1)B_{- \bf k}(t_1)\, .
\label{varia1}
\end{equation}
 \par
The time--translation invariance of the left--hand side of 
Eq.(\ref{varia1}) requires that the coefficients $B_{\bf k}(t)$ must 
be constant. Taking the Fourier transform, Eq.(\ref{varia1}) gives
\begin{eqnarray}
<\delta\varrho_{\bf k}(\omega)\delta\varrho_{-\bf k}(\omega^{\prime})>&=&
2\pi \delta(\omega + \omega^{\prime})\,
<(\delta\varrho_{\bf k}\delta\varrho_{-\bf k})(\omega)>
\nonumber
\\
&=&2\pi \delta(\omega + \omega^{\prime})\,D_{k}(\omega)
D_{k}(\omega^{\prime})\,|B_{ \bf k}|^2\, .
\label{varia2}
\end{eqnarray}
\par
By exploiting the fluctuation--dissipation theorem
\begin{equation}
<(\delta\varrho_{\bf k}\delta\varrho_{-\bf k})(\omega)>=\,
-\frac{2}{1-e^{-\beta\omega}}{\rm Im} D_k(\omega)\, ,
\label{fluct}
\end{equation}
where $\beta =1/T$ is the inverse temperature, (~we use units
such that $\hbar=~c=~k_B=1$~), we obtain for the coefficients 
$B_{\bf k}$ the equation
\begin{equation}
|B_{\bf k}|^2=\,
-\frac{2}{1-e^{-\beta\omega}}\,
\frac{{\rm Im} D_k(\omega)}{|D_k(\omega))|^2}\, .
\label{diff}
\end{equation}
We have used the relation $D_k(-\omega)=D^*_k(\omega)$.
Equation~(\ref{diff}) can be satisfied  only if the right--hand side
does not depend on $\omega$. This can occur only in particular 
situations, thus the original white--noise assumption about the
stochastic mean--field is not correct in general. This result is
quite general, so we can conclude that for a perturbed system
approaching  an equilibrium state, fluctuations about the
average trajectory cannot usually be accounted for by a 
white--noise stochastic force.
\par
Now, with reference to symmetric nuclear matter, we discuss 
particular physical situations in which Eq.(\ref{diff}) can have 
a solution. Only in such conditions the assumption of a withe--noise 
stochastic field is valid. The relevant quantity is the 
linear--response function $D_k(\omega)$. We evaluate $D_k(\omega)$ 
within a self--consistent mean--field approximation. In order to
derive compact analytical expressions, here we use the linearized 
Vlasov equation for calculating the response function. This 
equation can be regarded as a semiclassical approximation to 
the random phase approximation, valid in the longwavelength limit.
We also use a Skyrme--like form of the nucleon--nucleon effective 
interaction. Our self--consistent mean--field potential is given by
\begin{equation}
U=\,a\frac{\varrho}{\varrho_{eq}}+b\,\big({\frac{\varrho}{\varrho_{eq}}}\big)
^{\alpha+1}-d\,{\nabla}^2\varrho\, ,
\label{skyr}
\end{equation}
where $\varrho_{eq}$ is the saturation density of nuclear matter.
For the parameters in Eq.(\ref{skyr}) we take the
values:
\[
a=-356.8\,{\rm MeV},~~b=303.9\,{\rm MeV},~~\alpha=\,\frac{1}{6}~,
\]
\[
d=130\,{\rm MeV}\cdot {\rm fm}^5~.
\]
The values of $a$, $b$ and $\alpha$ reproduce the binding energy
(~$15.75\,{\rm MeV}$~) of
nuclear matter at saturation ($\varrho_{eq}=0.16\,{\rm fm}^{-3}$)
and give an incompressibility modulus of $201\,{\rm MeV}$.
For the values of $d$ we follow the prescriptions of
Ref.\cite{Mye66}.
\par
The response function is given by
\begin{equation}
D_k(\omega)=\,\frac{D_k^{(0)}(\omega)}
{ 1-{\cal A}_kD_k^{(0)}(\omega)}\, ,
\label{response}
\end{equation}
where $D_k^{(0)}(\omega)$ is the non--interacting particle--hole
propagator, and
\begin{equation}
{\cal A}_k=a\frac{1}{\varrho_{eq}}+\frac{b}{\alpha+1}\,\frac{1}
{\varrho_{eq}^{\alpha+1}}\varrho_0^{\alpha}+d\,k^2
\label{inter}
\end{equation}
are the Fourier coefficients of the effective interaction. Here
$\varrho_0$ is the density of the reference homogeneous state.

By substituting the expression (\ref{response}) for $D_k(\omega)$
into Eq.(\ref{diff}), we obtain
\begin{equation}
|B_{\bf k}|^2=\,
-\frac{2}{1-e^{-\beta\omega}}\,
\frac{{\rm Im} D_k^{(0)}(\omega)}{|D_k^{(0)}(\omega))|^2}\, .
\label{diff1}
\end{equation}
This equation shows that the coefficients $B_{\bf k}$ do not 
explicitely depend on the nucleon--nucleon effective interaction. 
However, we remark that $D_k^{(0)}(\omega)$ is the propagator of 
independent particles that are moving in the mean--field of the 
reference homogeneous state,
\begin{equation}
U_0=\,a\frac{\varrho_0}{\varrho_{eq}}+b\,\big({\frac{\varrho_0}
{\varrho_{eq}}}\big)^{\alpha+1}\, ,
\label{skyr0}
\end{equation}
thus the interaction between constituents does enter, although
not explicitly, into the expression of $B_{\bf k}$.
\par
We shall now show that in the classical limit $\omega/T\ll 1$,
the right--hand side of Eq.(\ref{diff1}) does not depend on $\omega$,
thus the assumption of a white--noise stochastic field can be
considered valid in that limit. \par
In the actual physical situations considered in this paper
the values of the temperature are small enough with respect to the
Fermi temperature so that the Pauli principle is still operating.
Therefore the strength of the particle--hole excitations
having energies much higher than $kv_F$ (~$v_F$ is the
Fermi velocity~) can be considered negligible. Moreover the
relevant values of the wave vector $k$ turn out to be such that
the quantity  $kv_F$ is of the same order of magnitude as $T$.
Thus the limit $\omega/T\ll 1$ also implies $\omega/kv_F\ll 1$.
\par
The non interacting particle--hole propagator $D_k^{(0)}(\omega)$
acquires a very simple form in the longwavelength  (Vlasov) limit.
The imaginary part is
\begin{equation}
{\rm Im} D_k^{(0)}(\omega)=\, \frac{1}{4\pi}\,m^2\frac{\omega}{k}\,
\int d\varepsilon_p\frac{\partial n_p}{\partial \varepsilon_p}\,
\theta\bigg(1-\frac{\omega}{kv}\bigg)\, ,
\label{imd0}
\end{equation}
where
\[
n_p=\,\frac{4}{e^{\beta(\epsilon_p-\tilde\mu)}+1}
\]
is the mean occupation number of nucleons with kinetic energy
$\epsilon_p=p^2/2m$, and $v=p/m$. The effective chemical potential 
$\tilde\mu$ is measured with respect to the uniform mean field $U_0$. 
\par
For $\omega/kv_F\ll 1$ the imaginary part of $D_k^{(0)}(\omega)$
is given by
\begin{equation}
{\rm Im} D_k^{(0)}(\omega)=\, -\frac{1}{\pi}\,m^2\frac{\omega}{k}\,
\frac{1}{e^{-\beta\tilde\mu}+1}+O\bigg((\frac{\omega}{kv_F})^3\bigg)\, ,
\label{imd01}
\end{equation}
while the real part of $D_k^{(0)}(\omega)$ in the longwavelength limit
takes the form
\begin{equation}
{\rm Re} D_k^{(0)}(\omega)=\, -\frac{1}{2\pi^2}\,
\int dp\,p^2\frac{\partial n_p}{\partial \varepsilon_p}\,
\bigg(-1+\frac{1}{2}\,\frac{\omega}{kv}\,
\ln\frac{1+\omega/kv}{|1-\omega/kv|}\,\bigg)\, .
\label{red0}
\end{equation}
For T sufficiently low with respect to $\tilde\mu$, the most
important contribution to the integral in Eq.(\ref{red0}) comes
from a small domain of $\varepsilon_p$ around $\tilde\mu$. So
we can take $\omega/kv\ll 1$ in evaluating the integral, and obtain
\begin{equation}
{\rm Re} D_k^{(0)}(\omega)=\,-\frac{\partial \varrho_0}{\partial \tilde\mu}
\,+O\bigg((\frac{\omega}{kv_F})^2\bigg)\, .
\label{red01}
\end{equation}
\par
With $D_k^{(0)}(\omega)$ given by Eqs.(\ref{imd01}) and (\ref{red01}),
the right--hand side of Eq.(\ref{diff1}) is independent of $\omega$ 
to the lowest significant order in $\omega/T$. Thus, for $\omega/T\ll 1$
the magnitude  of the coefficients $B_{\bf k}$ is given by
\begin{equation}
|B_{\bf k}|^2=\,\frac{2}{\pi}\,m^2(\frac{\partial \tilde\mu}
{\partial \varrho_0})^2\,\frac{T}{e^{-\beta\tilde\mu}+1}\frac{1}{k}\, .
\label{coeff}
\end{equation}
The phases of $B_{\bf k}$, instead, remain unknown. However, we will 
see that only the quantities $|B_{\bf k}|^2$ are needed to determine 
the probability distribution of density fluctuations. Finally, we
remark that $|B_{\bf k}|$ for a given ${\bf k}$, is determined
solely by the density and temperature of nuclear matter.
\par
The white--noise assumption is justified if the excitation strength
is concentrated in a narrow range of energy close to zero.
This condition requires that ${\rm Im} D_k(\omega)$ 
is a sharply peaked function in the proximity of
$\omega =0$ and is negligile elsewhere. The imaginary part of
$D_k(\omega)$ displays this feature for values of  temperature
and  density near the borders of the spinodal region, since the
pole of $D_k(\omega)$ lying on the imaginary axis, moves towards
$\omega =0$ as the system approaches the mechanical instability. 
This is shown in Fig.1, where we report ${\rm Im} D_k(\omega)$
calculated with Eq.(\ref{response}) using the complete
expression of $D_k^{(0)}(\omega)$. With our effective interaction,
for $T=5\,{\rm MeV}$ the spinodal region starts at $\varrho_{c}=0.617
\varrho_{eq}$. The values of $\varrho_0$ used in Fig.1 are close to
this critical value.

\subsection{Distribution of fluctuations.}

We now derive from Eq.(\ref{wiener}) the probability distribution 
for $\delta\varrho_{\bf k}(t)$ in the limit $\omega/T\ll 1$,
and for values of temperature and density in the proximity
of the spinodal region. The response function $D_k(\omega)$
has a pole in the lower part of the imaginary  $\omega$--axis,
at a position given by
\begin{equation}
i\Gamma_k=\,i\frac{\pi}{m^2}(1+e^{-\beta\tilde\mu})\frac{\partial\varrho_0}
{\partial\tilde\mu}\,\frac{\bigg({\displaystyle \frac{\partial^2 f}
{\partial\varrho_0^2}}|_T+dk^2\bigg)}
{{\cal A}_k}k\, .
\label{rate}
\end{equation}
We have used the relation
\begin{equation}
\frac{\partial\tilde\mu}{\partial\varrho_0}|_T=\,
\frac{\partial^2 f}{\partial\varrho_0^2}|_T-{\cal A}_0\, ,
\label{chem}
\end{equation}
where $f$ is the free--energy density and ${\cal A}_0={\cal A}_{k=0}$.
\par
In Eq.(\ref{rate}) the relevant quantity is the isothermal stiffness
${\displaystyle \frac{\partial^2 f}{\partial\varrho_0^2}}|_T$,
which vanishes on the
boundary of the spinodal region. Since we limit our calculations
to the proximity of the spinodal region, we neglect
${\displaystyle \frac{\partial^2 f}{\partial\varrho_0^2}}|_T$ with
respect to ${\cal A}_0$ in evaluating $D_k(t)$. Furthermore, in actual
calculations the typical values of $k$ which come into play
are such that the term $d\,k^2$ is  smaller than
$|{\cal A}_0|$, thus we also neglect this term with respect to
$|{\cal A}_0|$.  This approximation is consistent with
the longwavelength limit adopted in the
calculation of $D_k(\omega)$.
\par
Substituting into Eq.(\ref{wiener}) the response function
$D_k(t-t^\prime)$ calculated with these approximations,
the equation for the fluctuations $\delta\varrho_{\bf k}(t)$
becomes:
\begin{equation}
\delta\varrho_{\bf k}(t)=\,\delta\varrho_{\bf k}(t=0)e^{\Gamma_k t}+
\tilde B_k\,\int^t_0e^{\Gamma_k(t-t^\prime)}\,dW_k(t^\prime)\, ,
\label{ornul}
\end{equation}
where
\begin{equation}
|\tilde B_{k}|=\,\frac{1}{|{\cal A}_0|}\,\sqrt{\frac{2\pi T}
{m^2}\,(1+e^{-\beta\tilde\mu})\,k}\, ,
\label{difcoef}
\end{equation}
and $\Gamma_k$ is given by Eq.(\ref{rate}), neglecting the term
$d\,k^2$ in ${\cal A}_k$.
We recall that $\Gamma_k$ is negative, so that $|\Gamma_k|$
represents the damping rate of fluctuations, that 
vanishes for long wavelengths
when ${\displaystyle \frac{\partial^2 f}{\partial\varrho_0^2}}|_T
\rightarrow 0$.
\par
Equation (\ref{ornul}) represents an Ornstein--Uhlenbeck process
\cite{Gard} with $|\Gamma_k|$ as drift coefficient and $\tilde B_k$
as diffusion coefficient. The corresponding Fokker--Planck equation
for the probability distribution $P[\delta\varrho_{\bf k}(t)]$ reads
\begin{equation}
\frac{\partial}{\partial t}P[\delta\varrho_{\bf k}(t)]=\,|\Gamma_k|
\frac{\partial}{\partial\delta\varrho_{\bf k}(t)}\delta\varrho_{\bf k}(t)
P[\delta\varrho_{\bf k}(t)]+\frac{1}{2}\,
|\tilde B_k|^2\frac{\partial^2}{\partial\delta\varrho_{\bf k}^2(t)}
P[\delta\varrho_{\bf k}(t)]\, .
\label{fpe}
\end{equation}
\par
For simplicity we assume the state of the system at $t=0$ to be 
homogeneous on average (~$<\delta\varrho_{\bf k}(t=0)>=0$ for $k\not=0$~).
Equation (\ref{ornul}) says that this property holds during time
evolution. In this case the solution of Eq.(\ref{fpe}) is a
gaussian distribution with zero mean value. Whenever it is
necessary, a non vanishing mean value can easily be introduced.
The explicit expression of the distribution
$P[\delta\varrho_{\bf k}(t)]$ is
\begin{equation}
P[\delta\varrho_{\bf k}(t)]=\,N_1e^{\displaystyle{-\frac{1}{2}\sum_{\bf k}
\delta\varrho_{\bf k}^*(t)\frac{1}{\sigma^2_k(t)}
\delta\varrho_{\bf k}(t)}}\, ,
\label{gauss}
\end{equation}
with the variance $\sigma^2_k(t)$ given by
\begin{equation}
\sigma^2_k(t)=\,\sigma^2_k(t=0)\,e^{2\Gamma_k t}+\frac{T}
{f^{\prime\prime}+d\,k^2}\,
(1-e^{2\Gamma_k t})\, .
\label{variance}
\end{equation}
Here the constant $N_1$ is a normalization factor and we have introduced
the abbreviation
\[f^{\prime\prime}=\,\frac{\partial^2 f}{\partial\varrho_0^2}|_T\, .\]
For $t\rightarrow\infty$ Eq.(\ref{gauss}) reproduces the usual
gaussian approximation with variance
\begin{equation}
\sigma^2_k(t=\infty)=\,\frac{T}
{f^{\prime\prime}+d\,k^2}
\label{variance0}
\end{equation}
for the equilibrium thermodynamical fluctuations \cite{Landau}. We 
furthermore remark that Eq.(\ref{variance}) for the time evolution
of the variance is similar to that obtained
with different approaches in previous works on this subject
\cite{Col94,Kid95}. \par
For later purpose we report also the distribution of 
the fluctuations in ordinary space:
\begin{equation}
P[\delta\varrho({\bf r},t)]=\,N_2e^{\displaystyle {-\frac{1}{2}\int\,
d{\bf r}\,d{\bf r}^\prime\delta\varrho({\bf r},t)
M({\bf r},{\bf r}^\prime,t)\delta\varrho({\bf r}^\prime,t)}}\, ,
\label{gaussr}
\end{equation}
where
\[M({\bf r},{\bf r}^\prime,t)=\,\frac{1}{V}\sum_{\bf k}
e^{i{\bf k}\dot({\bf r}-{\bf r}^\prime)}\frac{1}
{\sigma^2_k(t)}\, ,\]
and $N_2$ is an appropriate normalization factor.
\par
The diffusion coefficients of Eq.(\ref{difcoef}) are derived by
means of the fluctuation--dissipation theorem, which concerns
only fluctuations about equilibrium.
In Ref.\cite{Hoff} a way has been suggested to extend
the treatment of stable cases to processes where
instabilities can develop.
Following that suggestion we include in our approach
the case of nuclear matter merged in the spinodal region.
In practice, we still assume the validity Eq.(\ref{gauss}) for the
probability distribution of the density fluctuations, with
the variance $\sigma_k^2(t)$ of Eq.(\ref{variance}) calculated
with the values of temperature and density of the new situation. 
This amounts to treating the diffusion coefficients for 
the unstable case as an analytic continuation of the stable--case 
coefficients in the $(\varrho,T)$ plane. The reliability of 
such a procedure lies in the fact that both the growing 
rate $\Gamma_k$ and the diffusion coefficient $\tilde B_k$ 
change smoothly when the system crosses the stability boundary 
and enters the spinodal region. The pole of $D_k(\omega)$
in turn, continuously moves along the imaginary axis from the
lower part to the upper part of the complex $\omega$--plane 
(~see Eq.(\ref{rate})~). In order to preserve causality, the integral 
for calculating the Fourier anti transform $D_k(t)$ must be 
performed along a path which cuts the imaginary axis above the pole.
\par
In the unstable case, the time behaviour of the variance 
$\sigma_k^2(t)$ in Eq.(\ref{variance}) is similar to that 
predicted by linear theories of the spinodal decomposition 
of alloys and fluids (~for an extensive review on this
subject see Ref.\cite{Gunt}~). The variance grows
exponentially for the fluctuations with wave number
\begin{equation}
k<k_c=\sqrt{\frac{|f^{\prime\prime}|}
{d}}\, ,
\label{kappac}
\end{equation}
while it tends to the aymptotic value $\sigma_k(t=\infty)$ of
Eq.(\ref{variance0}) for $k>k_c$. In particular
the growth rate $\Gamma_k$ presents a maximum for
$k=k_M=k_c/\sqrt{3}$. This means that the pattern of the regions 
which contain coherently correlated fluctuations is asymptotically 
characterized by the wavelength $\lambda_M=2\pi/k_M$.
These features for the growth rate of unstable modes are
analogous to those obtained in Ref.\cite{Col97} within a
different scheme.
\subsection{Size of fragments}
\par
Starting from the probability distribution for density 
fluctuations given by Eq.(\ref{gaussr}), we can determine the 
corresponding distribution for the size of the correlation domains. 
It has already been recalled that the stable and unstable cases 
can be treated within the same scheme. Thus we shall invesigate 
two different situations which could be explored by nuclear matter 
during a nucleus--nucleus collision: in one case the system 
is in the metastable region and relaxes towards a local
minimum of the free energy, while in the other case
the system is merged in the spinodal zone and develops 
density fluctuations which grow with time and will 
eventually lead to decomposition. According 
to our approximations, we limit our analysis
in both cases to values of temperature and density
in the proximity of the borders of the spinodal region. Moreover, 
we consider homogenous nuclear matter in both cases,
and still assume that $<\delta\varrho_{\bf k}(t=0)>=0$
for $k\not=0$.
\par
Before performing explicit calculations, we make a few remarks.
It is known that linear theories are unable to describe
the late stages of the spinodal decomposition of alloys and 
fluids (~see Ref.\cite{Gunt} and references quoted therein~). In
particular, they predict a limiting value for the length scale
that characterizes the pattern of the correlation domains.
This value is given by the wavelength $\lambda_M$ for which
the growth rate of fluctuations has a maximum.
Instead, Monte Carlo simulations and experimental results 
\cite{Leb82} show a continuous coarsening of the domains with 
increasing time. However it has been argued in Ref.\cite{Mar83} 
that the early--time Monte Carlo results are consistent with 
a linear theory, provided that a stochastic force is included.
\par
In the physical situations considered in the
present paper (heavy ion collisions), the value of the 
characteristic wavelength $\lambda_M$ is larger than
$10 \,{\rm fm}$, beyond the size of the nuclear system involved.
Moreover, the corresponding growth time $1/\Gamma_{k_M}$ is of
the same order of magnitude as the characteristic times of
the nucleus--nucleus collisions in the energy range considered 
here. Thus the fluctuations with wave number $k_M$ are still far
from being the predominant ones in this time interval. This means 
that the processes that we are investigating correspond to an 
early stage of the spinodal decomposition. Then we can expect 
reliable results from our approach, at least at a qualitative level.
\par
From Eq.(\ref{gaussr}) we obtain the usual expression for 
the equilibrium correlation function
\begin{equation}
G(|{\bf r}-{\bf r}^\prime|)=\,
\frac{1}{4\pi}\,\frac{T}{d}\,\frac{e^{-|{\bf r}-{\bf r}^\prime|/\xi}}
{|{\bf r}-{\bf r}^\prime|}\, ,
\label{corfeq}
\end{equation}
where
\[\xi=\,\sqrt{\frac{d}{f^{\prime\prime}}}\]
is the correlation length. This quantity, which  represents 
the average extension of the correlation domains, can be obtained 
by an appropriately weighted integral of the correlation function:
\begin{equation}
\xi=\,\int d{\bf r}d{\bf r}^\prime F({\bf r},{\bf r}^\prime)
G(|{\bf r}-{\bf r}^\prime|)\, .
\label{corleq}
\end{equation}
The function $ F({\bf r},{\bf r}^\prime)$ is a suitable weight
function. Here we extend this relation between 
averaged quantities to fluctuating quantities, for 
systems both at equilibrium and out of equilibrium. We 
then assume that the size of correlation domains at time $t$ 
is given by a quadratic functional of the fluctuations 
$\delta\varrho({\bf r},t)$:
\begin{equation}
b=\,L(t)\frac{\int d{\bf r}d{\bf r}^\prime
\delta\varrho({\bf r},t)F({\bf r},{\bf r}^\prime)
\delta\varrho({\bf r}^\prime,t)}
{\int d{\bf r}d{\bf r}^\prime F({\bf r},{\bf r}^\prime)
G(|{\bf r}-{\bf r}^\prime|,t)}\, ,
\label{size}
\end{equation}
where $L(t)=\, <b>$ is the length scale that characterizes
the pattern of the domains, and $G(|{\bf r}-{\bf r}^\prime|,t)$
is the correlation function for systems out of equilibrium.
The latter quantity is the space Fourier transform of the 
variance of Eq.(\ref{variance}).
\par
 In order to simplify calculations,
we further choose for $F({\bf r},{\bf r}^\prime)$ a separable form. 
The requirement that $b$ should be positive for any function
$\delta\varrho({\bf r},t)$, enforces a symmetric form
\begin{equation}
F({\bf r},{\bf r}^\prime)=\,f({\bf r})f({\bf r}^\prime)\, .
\label{funct}
\end{equation}
of the weight function. This form allows us to obtain 
a closed expression for the probability distribution of 
$b$. In addition, with this choice the final results are entirely 
independent of the function $f({\bf r})$.
\par
Now we derive the probability distribution for $b$
as a function of the length scale $L(t)$. Later we shall
give a procedure for determining $L(t)$.
\par
For a given probability distribution $P[\delta\varrho({\bf r},t)]$,
the related probability distribution for $b$, at a given 
time $t$, can be obtained by means of the functional integral 
\begin{equation}
P(b,t)=\,\int d[\delta\varrho({\bf r},t)]\,\delta\bigg(b
-\,\frac{L(t)}{C}\int d{\bf r}\delta\varrho({\bf r},t)F({\bf r})
\delta\varrho(0,t)\bigg)P[\delta\varrho({\bf r},t)]\, ,
\label{pb0}
\end{equation}
where we have put
\begin{equation}
C=\,\int d{\bf r}d{\bf r}^\prime f({\bf r})
G(|{\bf r}-{\bf r}^\prime|,t)f({\bf r}^\prime)
\label{const}
\end{equation}
in order to simplify the notation. \par
With the distribution of Eq.(\ref{gaussr}) and using the 
integral representation of the $\delta$--function
\[\delta(x)=\,\frac{1}{2\pi}\int d\eta e^{\displaystyle
i\eta x}\]
the equation for the distribution $P(b,t)$ takes the form
\begin{eqnarray}
P(b,t)&=&\,\frac{N_2}{2\pi}\int d\eta e^{\displaystyle
i\eta b}
\nonumber\\
& &
\label{pb}
\\
&\times&\int d[\delta\varrho({\bf r},t)]
e^{\displaystyle {-\frac{1}{2}\int\,
d{\bf r}\,d{\bf r}^\prime\delta\varrho({\bf r},t)\Big(
M({\bf r},{\bf r}^\prime,t)+2i\eta \frac{L(t)}{C}f({\bf r})
f({\bf r}^\prime)\Big)\delta\varrho({\bf r}^\prime,t)}}\, .
\nonumber
\end{eqnarray}
The functional integral is of gaussian type and allows us
to express the result of the integration in closed form:
\begin{equation}
P(b,t)=\,\frac{N_2}{2\pi}\int d\eta e^{\displaystyle
i\eta b}\frac{1}{(det\frac{1}{2\pi}[{\hat M}+
2i\eta \frac{L(t)}{C}{\hat F}])^{\frac{1}{2}}}\, .
\label{pdet}
\end{equation}
The quantities ${\hat M}$ and ${\hat F}$ are infinite--dimensional
operators, with matrix elements
$M({\bf r},{\bf r}^\prime,t)$ and $f({\bf r})f({\bf r}^\prime)$
respectively, in the coordinate representation.
\par
The determinant in the last equation can be factorized as
\begin{equation}
det\frac{1}{2\pi}[{\hat M}+2i\eta \frac{L(t)}{C}{\hat F}]=
det[\frac {\hat M}{2\pi}]
det[{\bf 1}+2i\eta \frac{L(t)}{C}{\hat M}^{-1}{\hat F}]\, ,
\label{facdet}
\end{equation}
where ${\bf 1}$ is the unit matrix. The square root of the first
factor on the  right--hand side and the normalization constant $N_2$ of
Eq.(\ref{pdet}) coincide and cancel.
What remains to be evaluated is the inverse of the square root
of the second determinant. For this purpose we write the determinant
in exponential form and expand the exponent in a power series.
Thus, we obtain the following formal expression
\begin{eqnarray}
(det[{\bf 1}+2i\eta \frac{L(t)}{C}{\hat M}^{-1}{\hat F}])^{-\frac{1}{2}}
&=&\,e^{\displaystyle -\frac{1}{2}\,
Tr\,\ln({\bf 1}+2i\eta\frac{L(t)}{C}{\hat M}^{-1}{\hat F})}
\nonumber\\
& &
\label{det}
\\
&=&\,e^{\displaystyle -\frac{1}{2}\sum\,\frac{1}{n}(-1)^{1+n}
(2i\eta \frac{L(t)}{C})^n\,Tr({\hat M}^{-1}{\hat F})^n}\, .
\nonumber
\end{eqnarray}
We recall that the matrix element $M^{-1}({\bf r},{\bf r}^\prime,t)$
and the correlation function
$G(|{\bf r}-{\bf r}^\prime|,t)$ coincide.
Because of the separable form chosen for the function
$F({\bf r},{\bf r}^\prime)$, Eq.(\ref{funct}), the trace operation
on the generic $n$--term of Eq.(\ref{det}) simply
yields $C^n$. Thus the series can be resummed and gives
$\ln(1+2i\eta L(t))$. Then the probability distribution $P(b,t)$ 
acquires the form
\[P(b,t)=\, \frac{1}{2\pi}\int d\eta \frac{ e^{\displaystyle
i\eta b}}{(1+2i\eta L(t))^{\frac{1}{2}}}\, .\]
A simple integration in the complex $\eta$  plane gives
the final result
\begin{equation}
P(b,t)=\,\frac{1}{\sqrt{\pi}}\frac{1}{\sqrt{2L(t)b}}
e^{\displaystyle -b/(2L(t))}\, .
\label{finpb}
\end{equation}
\par
From the probability distribution of the domain size we
can derive the distribution of the number of nucleons $A$ that are
contained in a correlation domain, assumed to be spherical.
For a homogeneous liquid
the relation between $A$ and the size $b$ is $b=\,2r_0 A^{1/3}$,
where $r_0$ is determined by the actual density. With a simple
transformation of variables we obtain for the probability
distribution of $A$, $P(A,t)$, the equation
\begin{equation}
P(A,t)=\,\frac{1}{3}\frac{1}{\sqrt{\pi}}\sqrt{\frac{r_0}{2L(t)}}\,
A^{-5/6}e^{\displaystyle -\frac{r_0}{L(t)}A^{1/3}}\, .
\label{distra}
\end{equation}
Further, to take into account that $A$ is a discrete variable
we express the probability of finding a correlation domain
containing $A$ nucleons, $Y(A)$, through the integral
\begin{equation}
Y(A)=\,\int _{A-1}^{A}dA\,P(A,t)\, .
\label{proba}
\end{equation}
For large $A$, $Y(A)$ tends to coincide with $P(A,t)$.

\section{Results}
\par
The distribution $P(A,t)$ and the probability $Y(A)$
are completely determined once the ratio
between the length scale $L(t)$ and the mean
interparticle spacing $r_0$ is fixed.
The parameter $L(t)$ sets the scale for the decrease of the 
correlation function  $G(r,t)$ with increasing $r$. 
We can obtain an estimate of $L(t)$ by analyzing the behaviour
of $G(r,t)$ as a function of $r$ at a given $t$. The correlation
function is initially determined by the variance
$\sigma^2_k(t=0)$, then, in the stable case, it asymptotically
assumes the form given in
Eq.(\ref{corfeq}), with the appropriate correlation length
$\xi=L(t=\infty)$, while in the unstable case, it acquires a 
damped oscillatory behaviour carachterized by the asymptotic 
wavelength $\lambda_M$. In order to illustrate the general features 
of the function $L(t)$, we simply assume that the initial 
fluctuations are negligible, $\sigma^2_k(t=0)\approx \,0$. In 
this case the function $G(r,t)$ is completely determined by the 
density and temperature of nuclear matter. Here we consider two 
sample values for the density (~$\varrho_0=\,0.65\varrho_{eq}$ and
$\varrho_0=0.58\varrho_{eq}$~)  and a single value for the
temperature  (~$T=5\,{\rm MeV}$~). This temperature is in the range
of values expected for the nuclear multifragmentation process 
\cite{Tam98}. The two corresponding points in the phase 
diagram $(\varrho,T)$ lie in the metastable region and in the 
spinodal region respectively, and are sufficiently close to 
the boundary of the spinodal zone to justify  our assumption
of a white--noise stochastic field. In Figs.~2 and 3 we show the
behaviour of $G(r,t)$ as a function of $r$ at three different 
values of time, both in the stable and unstable situations.
In the stable case of Fig. 2, a simple inspection of the behaviour
of $G(r,t)$ shows that it is reasonably well
reproduced by a function like that on the right--hand side of
Eq.(\ref{corfeq}) (obviously with $\xi$ replaced by $L(t)$~).
We adopt such a form for $G(r,t)$, then, by comparison
with its true behaviour shown in Fig.2, we can determine $L(t)$.
For the unstable case shown in Fig.~3, the situation is slightly 
more involved because the asymptotic regime is reached only after 
a very long time. For this case, we simply assume that $L(t)$ 
does coincide with the distance at which the value of $G(r,t)$ is 
reduced by $80\%$ with respect to its value at $r=1\,{\rm fm}$ 
(because of our approximations we cannot expect the present 
approach to be reliable for distances shorter than $1\,{\rm fm}$).
\par
At a given time $t$ the value of the length scale $L(t)$  
depends strongly on the distance from the boundary of the 
spinodal zone, the shorter this distance, the larger is $L(t)$.
In Fig.~4 the calculated length $L(t)$ is displayed as a function 
of $t$ for the two chosen sets of parameters. The values of $t$ 
are in the range that is relevant for nuclear fragmentation 
\cite{Tam98}. Figure 4 shows that for $t\sim 200\,{\rm fm}/c$, 
$L(t)$ pratically reaches its asymptotic value
($L(\infty)\approx 3.0\,{\rm fm}$) in the metastable situation, 
whereas in the unstable case $L(t)$ is still much smaller than
$L(\infty)\approx 12\,{\rm fm}$. \par
In the two physical situations considered here, two different
processes could drive nuclear matter towards a spinodal
decomposition. In the metastable case, if the density fluctuations
are large enough, the nuclear system can explore the unstable region
for a time sufficiently long to move towards a phase separation.
In the unstable case instead, fluctuations grow with time until they
cause the decomposition of the nuclear system. In both cases
we expect that the pattern of domains containing the liquid phase
is determined by the probability distribution $P(b,t)$ or $P(A,t)$
of Eqs.(\ref{finpb}) and (\ref{distra}).
\par
In order to assess the degree of validity of our approach, we
compare the results of our calculations with the corresponding
experimental data by identifying the probability $Y(A)$ of 
Eq.(\ref{proba}) with the distribution of the fragment yield.
Since experimentally the fragments are detected according to
their charge, we have to transform $P(A,t)$ and $Y(A)$ into
the corresponding functions of $Z$. We assume a homogeneous
distribution also for the charge
$Z=\,{\displaystyle \frac{(1-\alpha)}{2}}A$, with $\alpha=\,(N-Z)/A$,
and use $\alpha=\,0.2$, which
 corresponds to the average asymmetry of the nuclear
systems considered.
\par
In Fig. 5 the probability $Y(Z)$ is displayed as a function of $Z$
on a double logarithmic scale for three different values 
of the ratio $L(t)/r_0$.
The range of values for $L(t)/r_0$ has been chosen in accordance
with that of $L(t)$ in Fig. 4. Figure 5 shows that $Y(Z)$ can
be fit with good accuracy by a power law $Y(Z)=Y_0Z^{-\tau_{eff}}$.
The values of the effective exponent, $\tau_{eff}$, lie
between $1.17$ for $L(t)/r_0=\,4$ and $1.42$ for $L(t)/r_0=\,2$.
\par
The power--law behaviour of the fragment yield and the determination
of the exponent have been the subject of several experimental
studies of multifragmentation (~see for example the recent
papers \cite{T.Li94,Dag95}~). The observed values of the exponent are
in the interval $\sim 1.2-1.5$ for nuclear rections with
beam energies lower than $\sim 40\,A\,{\rm MeV}$, whereas 
they exceed the value of $2$ at higher energies \cite{T.Li94,Dag95}. 
A value of the exponent  $\tau_{eff}\geq 2$ can be unlikely 
reproduced by our calculations because we would need an unreasonably 
low value for the ratio $L(t)/r_0$. However, in various papers 
\cite{Dag95,Hsi95,Kun95} it has been remarked that the effects of 
collective motions, that have not been taken into account by our 
present approach, should become more important with 
increasing beam-energy.
\par
Figures~6 and 7 show a comparison between the charge
distributions predicted by our approach, $Y(Z)$, and  recent
experimental data obtained by the Multics/Miniball collaboration for
$Au+Au$ collisions at an incident energy of $E= 35~A\,{\rm MeV}$ 
\cite{Dag98} and by the INDRA Collaboration for $^{129}Xe+Sn$ and
$^{155}Gd+^{238}U$ collisions at $E=32~A\,{\rm MeV}$ and 
$E=36~A\,{\rm MeV}$ respectively \cite{Riv98}. The calculations 
have been performed for three values of the parameter $L(t)/r_0$. 
We have normalized the experimental distributions 
to one in order to perform the comparison on an absolute scale. 
We can see that the agreement between experimental 
data and the calculated charge distributions is quite 
satisfactory for $Z<30\div 35$ and that for the lighter 
fragments the experimental points are better reproduced with 
larger values of the ratio $L(t)/r_0$. For $Z>30\div 35$ 
the observed distribution presents a slope steeper than that 
predicted by our calculations. This faster decrease should be 
ascribed to finite--size effects \cite{Dag96} which have not
been included in our nuclear matter treatment.

\section{Summary and conclusions}

We have studied the density fluctuations associated with a
one--body treatment of nuclear dynamics. In our approach the 
fluctuations are generated by adding a stochastic term to the 
mean field. This additional random force is determined by a 
self-consistency condition required by the fluctuation--dissipation 
theorem. We have treated the effects of the stochastic
field in linear approximation and this has allowed us to express 
the time evolution of the fluctuations in a closed form.

First we have analyzed the nature of the stochastic field and have 
shown that in general a white--noise assumption for the stochastic 
field is not consistent with the fluctuation--dissipation theorem. 
Then we have studied the particular physical conditions in which
the white--noise nature of the stochastic term can be retained.
These conditions include hot nuclear matter at a temperature
$T\approx5~{\rm MeV}$, where the system can be still considered 
degenerate. We have found that for a Fermi system the treatment 
of density fluctuations by means of a white--noise stochastic term 
is justified when the limit $\omega/T\ll 1$ gives
a reasonable approximation to  the density--density response.
This condition is better satisfied when the density
and temperature of the system are close to the borders
of the spinodal region in the ($\varrho,T$) plane. Thus, in the
limit $\omega/T\ll 1$ the equilibrium fluctuations can be adequately
described  by means of thermodynamic functions and we can 
expect that in this limit the purely quantum fluctuations 
will play a negligible role also for systems not too far from 
equilibrium. We have extended the results obtained for the 
probability distribution of a metastable system to unstable 
situations. This has been achieved by extrapolating the relevant 
quantities across the boundary of the spinodal region. Because 
of the linear approximation used for evaluating the response 
of the system to the stochastic force, the fluctuations have 
a gaussian probability distribution.
\par
In the final part of this paper we have introduced a procedure
to determine the size and mass distributions of the domains
containing correlated density fluctuations, then we have 
compared the obtained mass distribution to the yield of light 
fragments observed in the multifragmentation of heavy nuclei.
The procedure  proposed here is quite general and can
be applied to any gaussian fluctuation distribution
\par
Our approach can account both for the observed power--law 
distribution and for the value of the effective exponent 
found experimentally, but for the exponent the agreement is
limited to collisions with beam energies lower than 
$\sim 40~A\,{\rm MeV}$. This discrepancy between our predictions
and the observed values of the effective exponent in collisions
of higher energies deserves further investigations. A detailed
comparison with experiment has shown that our approach
fairly reproduces the measured charge distributions for
$Z<30\div 35$. Since we are dealing with infinite nuclear matter, 
we expect to overestimate the number of fragments having a large 
fraction of the mass of the emitting source.
\par
Finally, we remark that the obtained mass distribution
contains only one parameter, the ratio between the time--dependent
length scale of  domains $L(t)$ and the mean interparticle
spacing $r_{0}$. This ratio can become large. 
A more detailed comparison of the present model with experimental
data could also give an estimate of the time  required by the system
to break up.

\newpage

Figure captions:
\vskip 0.5cm
\begin{description}
\item{Fig.1~}Imaginary part of the response function $D_k(\omega)$
of Eq.(\ref{response}) for hot nuclear matter ($T=5\,{\rm MeV}$)
at different densities approaching the
critical value $\varrho_{c}=0.617\,\varrho_{eq}$ 
(dotted line: $\varrho_{0}=0.70\,\varrho_{eq}$, 
dashed line: $\varrho_{0}=0.65\,\varrho_{eq}$,
full line: $\varrho_{0}=0.63\,\varrho_{eq}$). The value of $k$ is
$0.1\,k_{F}$.
\bigskip

\item{Fig.2~}Spatial behaviour of time--dependent correlation
function $G(r,t)$ for nuclear matter on the stable side of the
spinodal curve ($\varrho_{0}=0.65\,\varrho_{eq}$, $T=5\,{\rm MeV}$).
The three curves correspond to different values of $t$ (full line: 
$t=50\,{\rm fm}/c$, dashed line: $t=100\,{\rm fm}/c$, dotted line:
$t=200\,{\rm fm}/c$).
\bigskip
\item{Fig.3~}Same as Fig.2, but for the unstable case
($\varrho_{0}=0.58\,\varrho_{eq}$).

\bigskip
\item{Fig.4~}Behaviour of the length scale $L(t)$ within the spinodal
region (full line: $\varrho_{0}=0.58\,\varrho_{eq}$) and outside it 
(dashed line: $\varrho_{0}=0.65\,\varrho_{eq}$). 

\bigskip
\item{Fig.5~}Fragment distribution $Y(Z)$ calculated for different 
values of the ratio $L(t)/r_{0}$. Full line: $L(t)/r_{0} =4$, 
dashed line: $L(t)/r_{0}=3$, dotted line: $L(t)/r_{0}=2$.

\bigskip
\item{Fig.6~}Comparison of fragment distribution $Y(Z)$ calculated
for $L(t)/r_{0}=6$ (full line), $4$ (dashed line), $2$ (dotted line) 
with experimental distribution for the reaction $Au+Au$ at $35~A\,
{\rm MeV}$. Data from Ref.\cite{Dag98} have been normalized to one. 

\bigskip
\item{Fig.7~}Same as Fig.6, but for the reactions $^{129}Xe+Sn$ at
$E=32~A\,{\rm MeV}$ (triangles) and
$^{155}Gd+^{238}U$ at $E=36~A\,{\rm MeV}$ (circles). Data from Ref.
\cite{Riv98} have been normalized to one. 
\end{description}
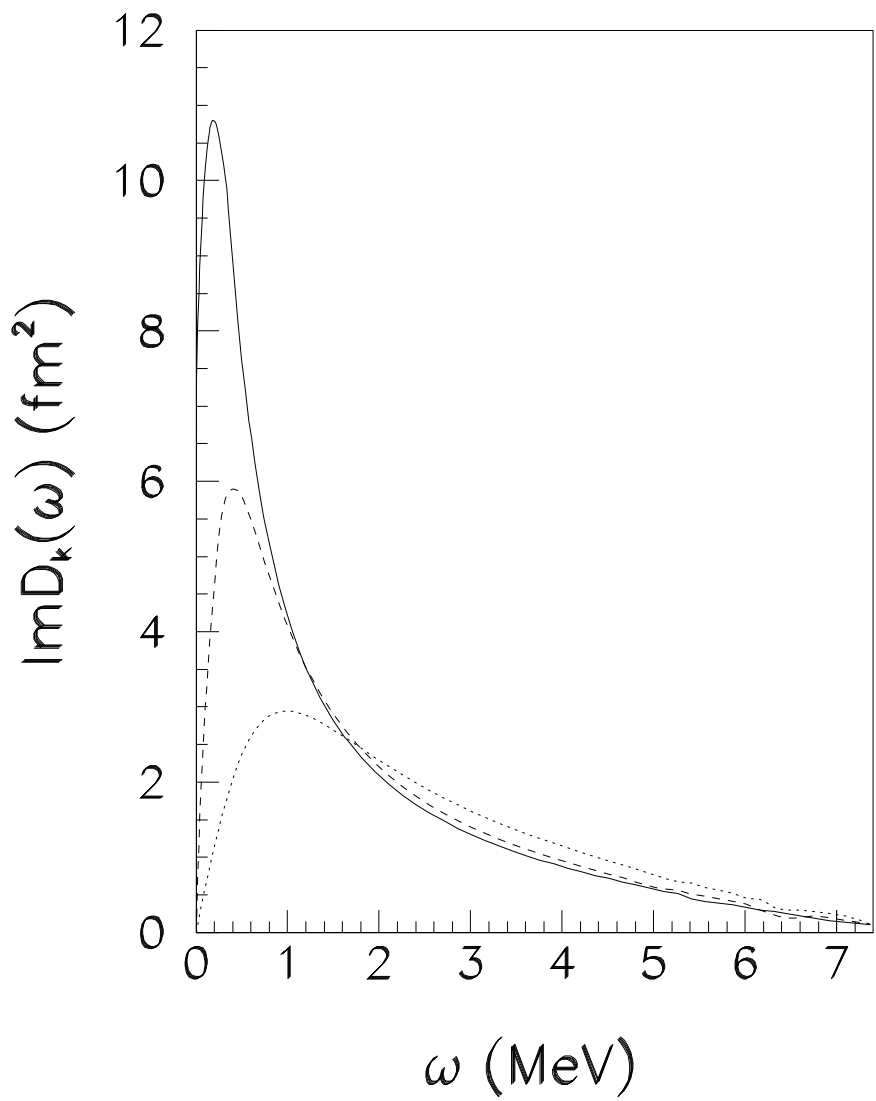
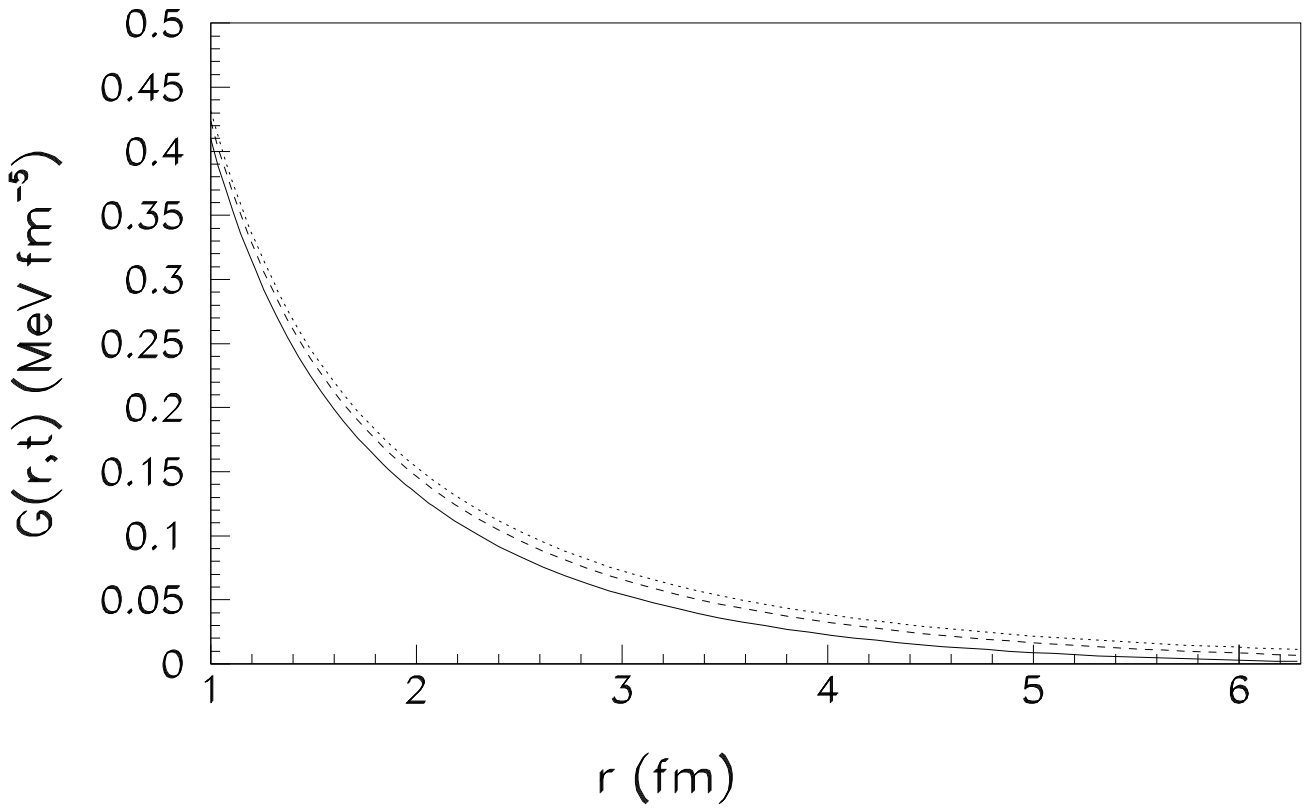
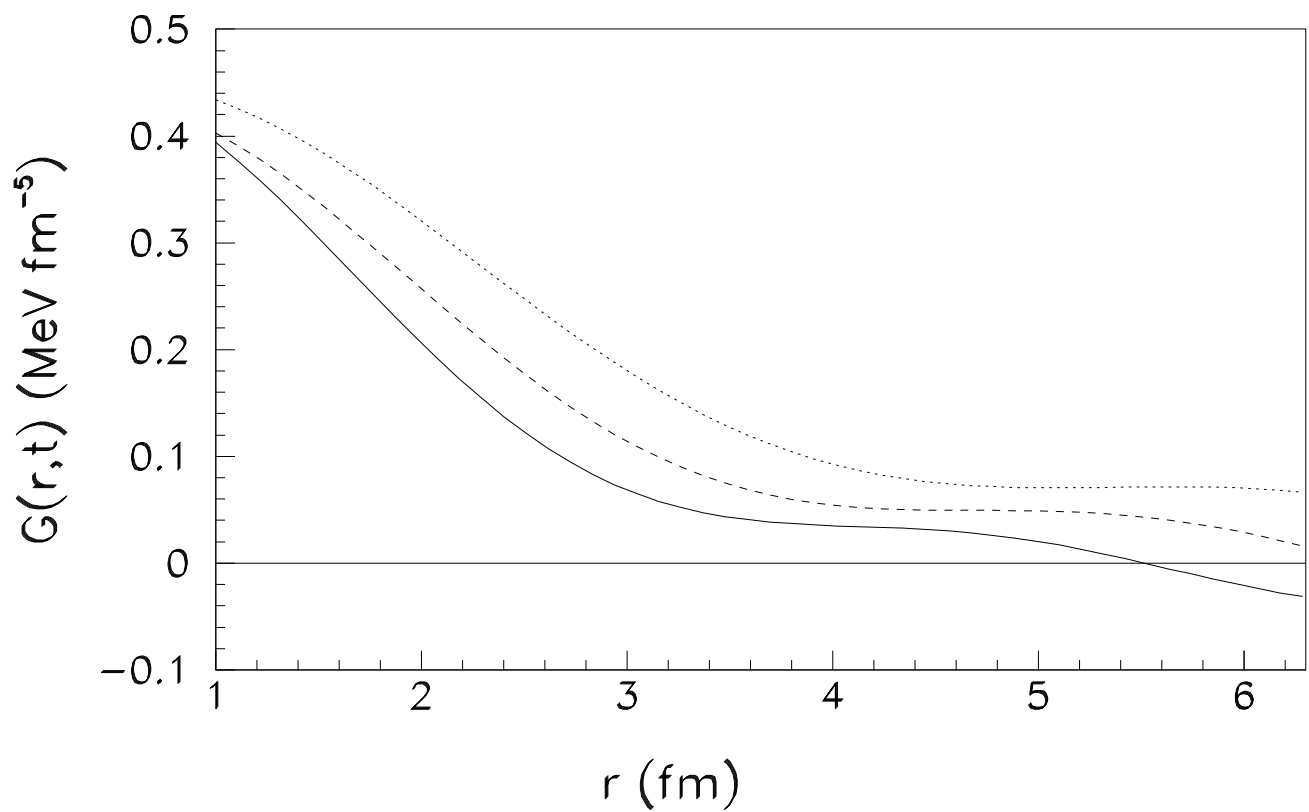
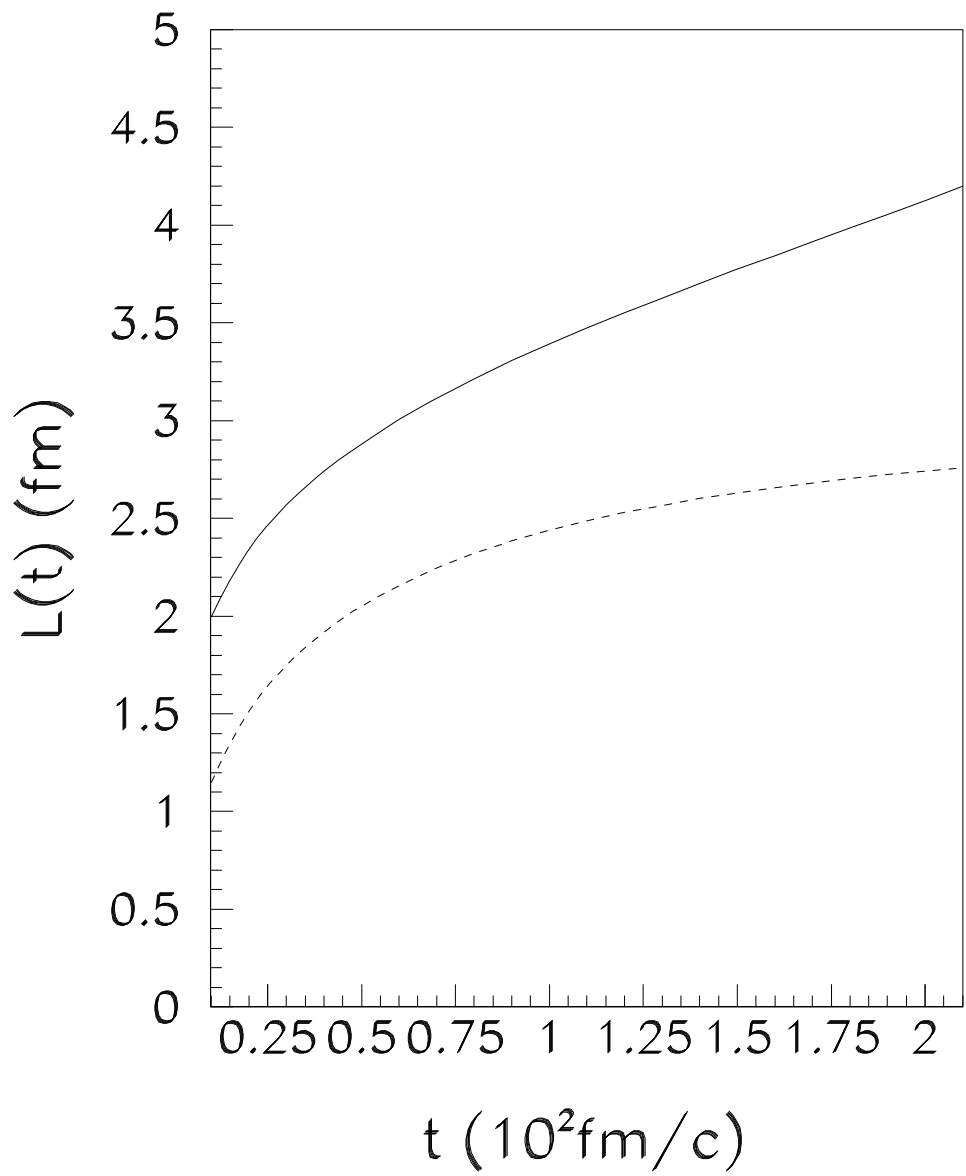
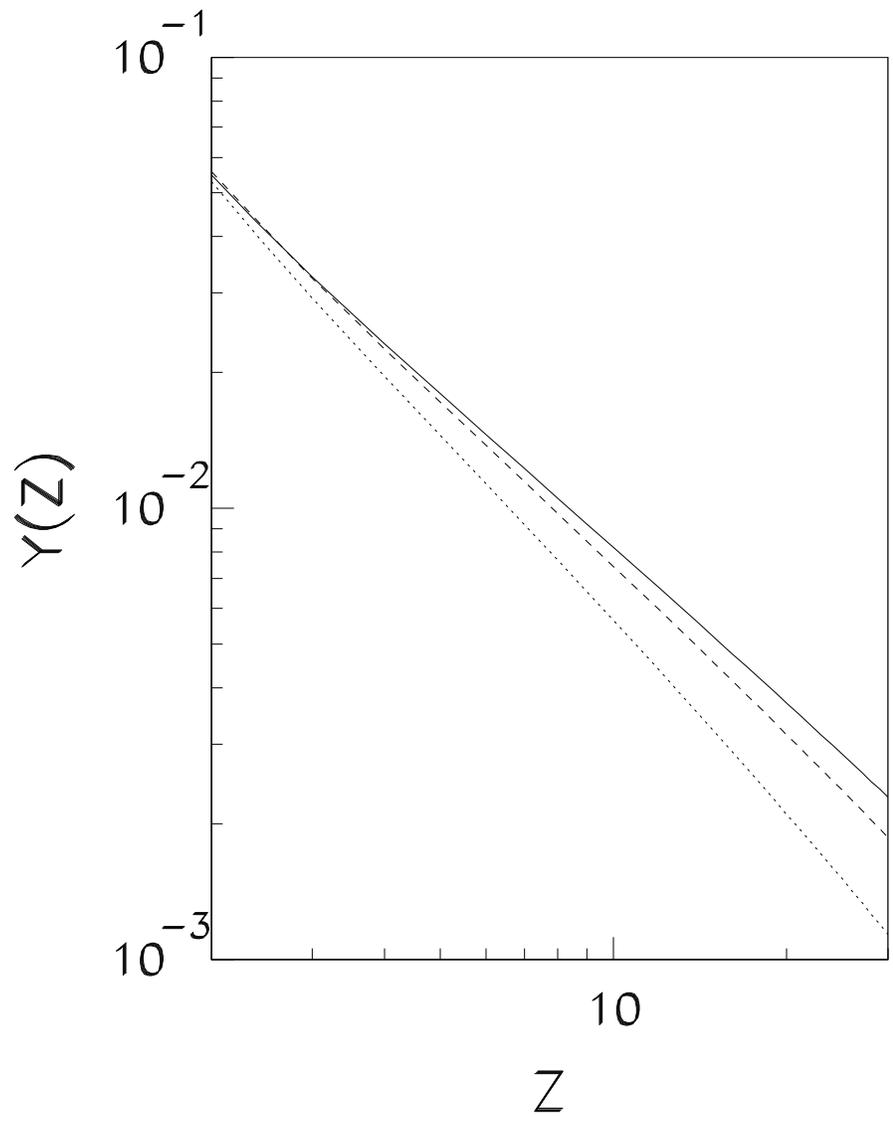
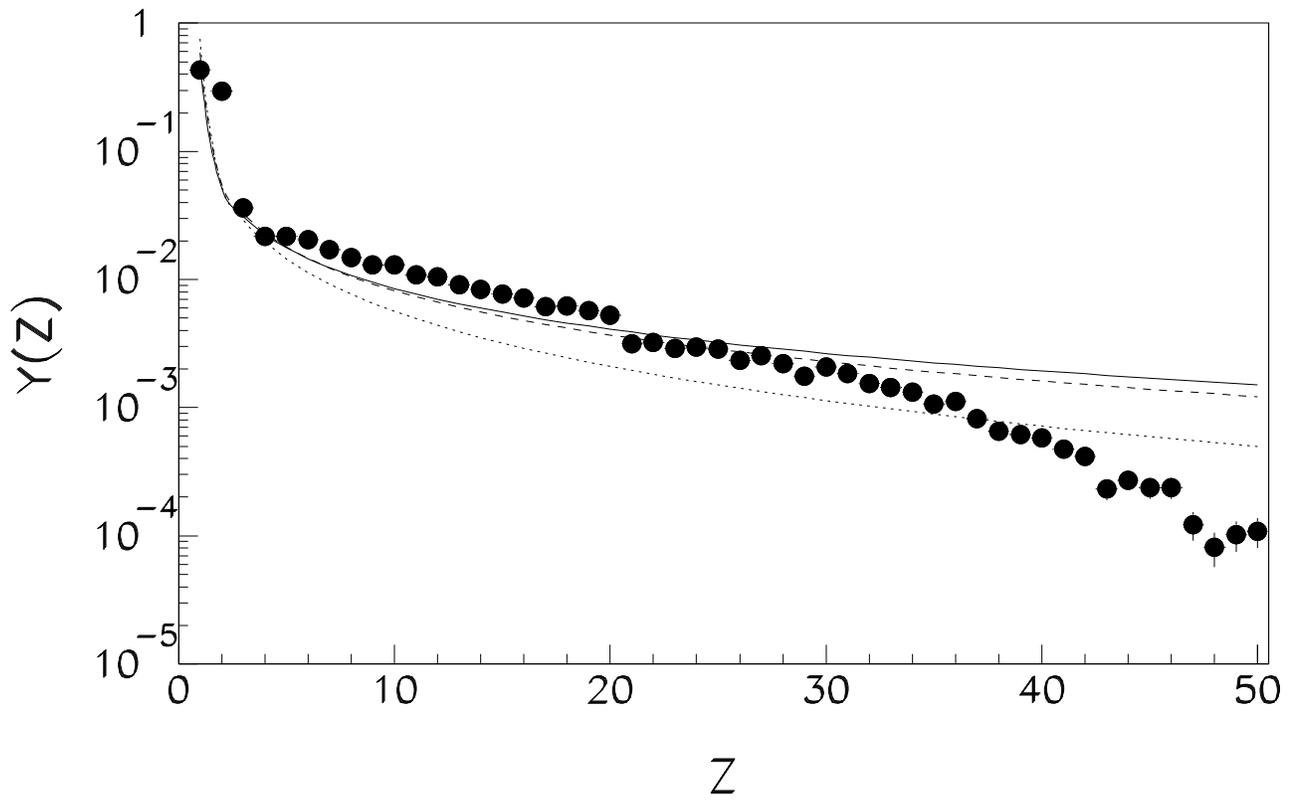
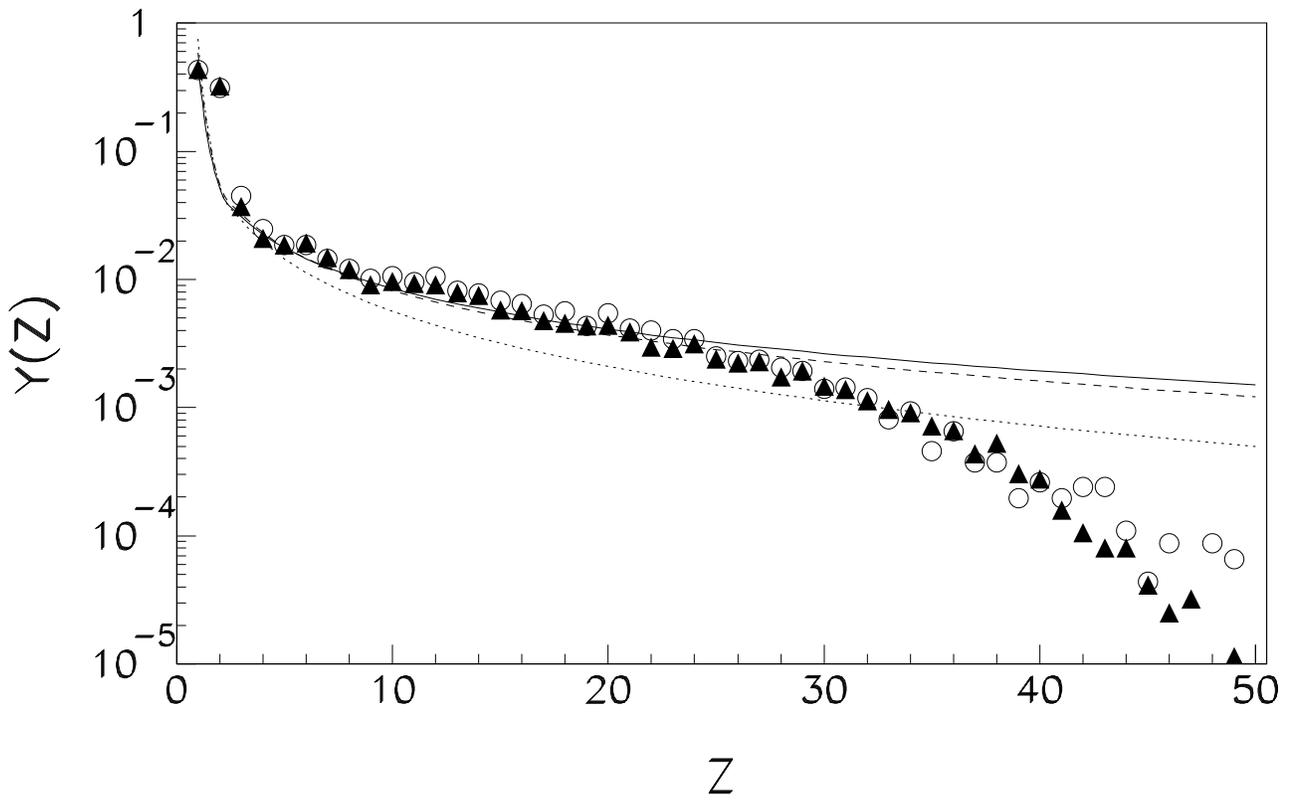
\end{document}